\documentclass[letter]{aa} 

\usepackage{graphicx}
\usepackage{txfonts}
\usepackage[citecolor=blue, linkcolor=blue, urlcolor = black, colorlinks = true, backref = page]{hyperref}
\usepackage{ulem}
\usepackage{xcolor}

\begin{document} 

   \title{Asymmetries of frequency splittings of dipolar mixed modes:\\
   a window on the topology of deep magnetic fields}

   \author{S. Mathis
          \inst{1}, L. Bugnet \inst{2}
          }

   \institute{Université Paris-Saclay, Université Paris Cité, CEA, CNRS, AIM, 91191, Gif-sur-Yvette, France
    \email{stephane.mathis@cea.fr}\\
   \and Institute of Science and Technology Austria (IST Austria), Am Campus 1, Klosterneuburg, Austria}

   \date{Received XX; accepted YY}

\titlerunning{Frequency splitting asymmetries and deep magnetic field topology}

 
  \abstract
   {Space asteroseismology is revolutionizing our knowledge of the internal structure and dynamics of stars. A breakthrough is ongoing with the recent discoveries of signatures of strong magnetic fields in the core of red giant stars. The key signature for such a detection is the asymmetry these fields induce in the frequency splittings of observed dipolar mixed gravito-acoustic modes.}
   {We investigate the ability of the observed asymmetries of the frequency splittings of dipolar mixed modes to constrain the geometrical properties of deep magnetic fields.}
   {We use the powerful analytical Racah-Wigner algebra used in Quantum Mechanics to characterize the geometrical couplings of dipolar mixed oscillation modes with various possible realistic fossil magnetic fields' topologies and compute the induced perturbation of their frequencies.}
   {First, in the case of an oblique magnetic dipole, we provide the exact analytical expression of the asymmetry as a function of the angle between the rotation and magnetic axes. Its value provides a direct measure of this angle. Second, considering a combination of axisymmetric dipolar and quadrupolar fields, we show how the asymmetry is blind to unravel the relative strength and sign of each component. Finally, in the case of a given multipole, we show that a negative asymmetry is a signature of non-axisymmetric topologies.}
   {Asymmetries of dipolar mixed modes provide key but only partial information on the geometrical topology of deep fossil magnetic fields. Asteroseismic constraints should therefore be combined with spectropolarimetric observations and numerical simulations, which aim to predict the more probable stable large-scale geometries.}


   \maketitle
%
\section{Introduction}

For the past two decades, space asteroseismology has made the dream of Sir Eddington to "see" the interiors of stars a reality. Beginning with Helioseismology for our Sun and moving to the stars of our galaxy with asteroseismology, frequencies of Solar and stellar oscillation modes have progressively allowed us to constrain the internal solar and stellar structure and dynamics with a precision that has never been reached before. First, stellar seismology allowed us to constrain the hydrostatic and thermodynamic equilibrium structures of stars at different evolutionary stages \citep[e.g][]{HekkerJCD2017,JCD2021,Aerts2021}. Next, it demonstrated that stellar interiors are the seats of an extremely intense transport of angular momentum for all stellar masses at all ages \citep[][and references therein]{Garciaetal2007,Aertsetal2019}. Finally, it opened the unique possible window to constrain deep internal magnetic fields \citep[e.g.][]{GoodeThompson1992,Hasanetal2005,Bugnetetal2021,Mathisetal2021} and magnetic activity \citep[e.g.][]{Santosetal2021}.

In this context, the recent discovery of deep magnetic fields in the core of three red giant stars by \cite{Lietal2022}, followed by ten more magnetized red giants reported in \cite{Deheuvelsetal2023}, constitutes an extraordinary breakthrough. Indeed, this is the first time we obtain constraints on possible fossil fields in the stably stratified core of low- and intermediate-mass evolved stars. These fields result from the relaxation towards an equilibrium configuration of past dynamo magnetic fields \citep[e.g.][]{BraithwaiteSpruit2004,DuezMathis2010}, for instance those generated in the convective core of main-sequence stars with masses above $\sim$ 1.1 M$_{\odot}$ \citep[we refer the reader to Fig. 1 in][which details this formation mechanism]{Bugnetetal2021}. This discovery is of major importance since these fields are one of the possible mechanisms to explain the observed strong transport of angular momentum in stellar interiors \citep[e.g.][]{Spruit1999,Spruit2002,MathisZahn2005,Fulleretal2019,Eggenbergeretal2022,Petitdemangeetal2023}. Moreover, before this discovery, such fossil fields have only been observed at the surface of $\sim 10\%$ early O,B,A-type stars thanks to high-precision spectropolarimetry \citep{Wadeetal2016} but never in the deep interiors of stars. This discovery strengthens the role of red giant stars as one of the cornerstones of modern stellar astrophysics. Indeed, these stars have also allowed during the last decade to improve our knowledge of stellar evolutionary stages \citep{Beddingetal2011} and of the core to the surface rotation contrast \citep[e.g.][]{Mosseretal2012,Deheuvelsetal2014,Gehanetal2018}, and thus of the internal transport of angular momentum. 

The key to these major advances is the coupling between acoustic and gravity modes. Mixed gravito-acoustic oscillation modes behave as acoustic modes in the deep convective envelope of red giant stars and as gravity modes in their radiative core \citep[e.g.][]{Shibahashi1979}. Therefore, they allow us to probe these stars near the surface and in the core. In the case of deep magnetic fields, the key parameter is the splitting of the frequencies of mixed modes due to the combined action of rotation and magnetism. In the case of magnetism, the Lorentz force causes an asymmetric frequency splitting as this has been shown by \cite{Bugnetetal2021} and \cite{Bugnet2022} in the case of axisymmetric dipolar fields and by \cite{Lietal2022} in the case of a general 3D magnetic topology. If the asymmetry in the power spectrum density provides constraint on the amplitude of the internal magnetic field, the relation between the asymmetry and the 3D topology of the magnetic field as presented in \cite{Lietal2022} might be degenerate.

The objective of our study is therefore to explore in more detail the relation between the asymmetry parameter $a$ as defined in \cite{Lietal2022} and the magnetic field topology. We first derive analytically in Sec. \ref{sec:Dipolar} the expression of $a$ for a general magnetic topology by using the Racah-wigner algebra used in Quantum Mechanics. This allows us to explore in Sec. \ref{sec:topology} its behavior for a set of chosen magnetic topologies based on spectropolarimetric observations of fossil fields and on the state-of-the-art of our theoretical knowledge on the formation and the stability of these fields. The results we obtain allow us to quantify the strength and weaknesses of $a$ as a window on the topology of deep magnetic fields and to conclude on the crucial importance of continuing studies to predict fossil fields' magnetic topologies and novel inversion techniques to overrule its potential degeneracy. 

\section{Mixed modes perturbations by the combined action of magnetism and rotation}

\subsection{General perturbative formalism}
\label{sec:General}
We consider an evolved low-mass or intermediate-mass star where mixed gravito-acoustic modes propagate, and we assume that its central stably stratified radiative core is the seat of a deep steady\footnote{i.e. with a characteristic evolution time longer than the rotation and the pulsation periods.} magnetic field. Its radial component is expanded on spherical harmonics defined in Eq.~(\ref{eq:defSH}): 
\begin{equation}
B_r\left(r,\theta,\varphi\right)=\sum_{l=1}^{L_B}\sum_{m=-l}^{l} b^{l}_{m}\left(r\right)Y_{l}^{m}\left(\theta,\varphi\right),
\label{eq:Br}
\end{equation}
where $r$, $\theta$, and $\varphi$ are the radius, co-latitude, and azimuth, respectively, and $L_B$ is the formal limit of the expansion. Following \cite{Goupiletal2013}, we introduce $\langle\Omega\rangle_{\rm g}$ (resp. $\langle\Omega\rangle_{\rm p}$) the averaged rotation of the cavity where mixed modes behave as gravity (resp. acoustic) modes.

In the case of non-rotating non-magnetic stars, an oscillation eigenmode is purely spheroidal and can be expanded on vectorial spherical harmonics \citep{Rieutord1987}:
\begin{eqnarray}
\lefteqn{{\boldsymbol\xi}_{0;n,l,m}\left(r,\theta,\varphi\right)=\left[{\xi}_{r;n,l}\left(r\right){\boldsymbol R}_{l}^{m}\left(\theta,\varphi\right)+{\xi}_{h;n,l}\left(r\right){\boldsymbol S}_{l}^{m}\left(\theta,\varphi\right)\right]}\nonumber\\
&&\times\exp(-i\omega_{0;n,l}t),
\end{eqnarray}
where ${\boldsymbol R}_{l}^{m}\left(\theta,\varphi\right)=Y_{l}^{m}\left(\theta,\varphi\right){\bf e}_{r}$ and ${\boldsymbol S}_{l}^{m}\left(\theta,\varphi\right)={\boldsymbol\nabla}_{\rm H}Y_{l}^{m}\left(\theta,\varphi\right)$ with ${\boldsymbol\nabla}_{\rm H}={\bf e}_{\theta}\partial_{\theta}+1/\sin\theta\,{\bf e}_{\varphi}\partial_{\varphi}$, $\left\{e_j\right\}_{j\equiv r,\theta,\varphi}$ being the classical spherical unit-vector basis. The unperturbed mixed mode frequency $\omega_{0;n,l}$ is the eigenvalue of the linearized oscillations equations which can be written formally as  
\begin{equation}
{\mathcal L}_{0}{\boldsymbol\xi}_{0;n,l,m}=\omega_{0;n,l}^{2}\,{\boldsymbol\xi}_{0;n,l,m},
\label{EigenO}
\end{equation}
where ${\mathcal L}_{0}$ is a linear operator operating on ${\boldsymbol\xi}_{0;n,l,m}$, which accounts for the compressibility, the buoyancy, and the self-gravity of the mode; we refer the reader to \cite{Unnoetal1989} for its detailed expression. Because of the degeneracy with $m$ of Eq. (\ref{EigenO}), any linear combination of eigenmodes: 
\begin{equation}
{\boldsymbol\xi}_{0;n,l}=\sum_{m=-l}^{l} a_{m}{\boldsymbol\xi}_{0;n,l,m} 
\end{equation}
is an eigenfunction of ${\mathcal L}_{0}$ associated with $\omega_{0;n,l}$.

A deep magnetic field as those detected in \cite{Lietal2022} and \cite{Deheuvelsetal2023} and moderate rotation trigger first-order perturbations of the eigenfrequencies (eigenfunctions) such that $\omega_{n,l,m}=\omega_{0;n,l}+\omega_{1;n,l,m}$ with $\omega_{1;n,l,m}\!<\!\!<\!\omega_{0;n,l}$ (respectively ${\boldsymbol \xi}_{n,l,m}={\boldsymbol \xi}_{0;n,l}+{\boldsymbol\xi}_{1;n,l,m}$ with $\vert\vert{\boldsymbol\xi}_{1;n,l,m}\vert\vert\!<\!\!<\!\vert\vert{\boldsymbol \xi}_{0;n,l}\vert\vert$). In the frame rotating with $\langle \Omega\rangle_{\rm g}$, they can be computed by solving the linear system 
\begin{equation}
    \omega_{1}{\boldsymbol a}=(\zeta{\boldsymbol M} + {\boldsymbol R}){\boldsymbol a}.
\label{eq:system}
\end{equation}
The diagonal ${\boldsymbol R}$ matrix is associated to the Coriolis acceleration and the residual core-envelope differential rotation in the frame rotating with the g-dominated modes propagation cavity. The matrix ${\boldsymbol M}$ is associated to the Lorentz force and its elements at a fixed $l$ are given by
\begin{equation}
    M_{m,m'}=
    \frac{\left<{\boldsymbol\xi}_{0;n,l,m},{\mathcal L}_{L}\left[{\boldsymbol\xi}_{0;n,l,m'}\right]\right>}{2\omega_{0;n,l}I_{n,l}},
\label{eq:MatrixLorentz}
\end{equation}
where
\begin{eqnarray}
    \lefteqn{{\mathcal L}_{\rm L}\left[{\boldsymbol\xi}\right]
    =  -\frac{1}{\rho\mu_0}\left[({\boldsymbol\nabla} \times {\boldsymbol B'}) \times {\boldsymbol B_0} + ({\boldsymbol\nabla} \times {\boldsymbol B_0}) \times {\boldsymbol B'}\right]}\nonumber\\ 
    &&- \frac{{\boldsymbol \nabla}\cdot\left(\rho{\boldsymbol\xi}\right)}{\rho^2\mu_0}\left[{\boldsymbol\nabla}\times{\boldsymbol B_0}) \times {\boldsymbol B_0}\right],
\label{LorentzOperator}    
\end{eqnarray}
with the background magnetic field ${\boldsymbol B}_0$, the hydrostatic density profile of the star $\rho$, the Eulerian perturbation of the magnetic field induced by the oscillation ${\boldsymbol B}' = {\boldsymbol\nabla} \times ({\boldsymbol\xi} \times {\boldsymbol B}_0)$, and the magnetic permeability of vacuum $\mu_0$. We neglect here the indirect terms that are induced by the deformation of the hydrostatic structure of the star by the Lorentz force \citep{GoughThompson1990}. This is justified in Appendix \ref{Appen:Lorentz} for the case of mixed modes propagating in the core of red giant stars. We have introduced the inner product $\left<f,g\right>=\int f^{*}\,g\,{\rho}\,r^2 {\rm d}r\, \sin\theta\,{\rm d}\theta\,{\rm d}\varphi$. Finally, $I$ is the mode inertia defined as:
\begin{equation}
    I_{n,l}=\left<{\boldsymbol\xi}_{0;n,l,m},{\boldsymbol\xi}_{0;n,l,m}\right>= \int_{0}^{R}(\vert\xi_{r;n,l}\vert^2+l(l+1) \vert\xi_{h;n,l}\vert^2) \rho r^2 \,\mathrm{d}r,
\end{equation}
where $R$ is the radius of the star.

In the region where \cite{Lietal2022} and \cite{Deheuvelsetal2023} have detected and probed deep magnetic fields, mixed modes behave as asymptotic g-modes. This allows us to simplify the previous equations. First, in this regime, the acoustic behaviour of the mode is weak and can be filtered out. That allows us to neglect the third term in Eq. (\ref{LorentzOperator}) because of the corresponding anelastic behaviour of the modes where ${\boldsymbol\nabla}\cdot\left(\rho{\boldsymbol\xi}\right)\approx 0$. In this regime, modes are rapidly oscillating along the radial direction (i.e. $k_{r;n,l}\!>\!>\!k_{h;l}$, where $k_{r;n,l}$ and $k_{h;l}$ are the local vertical and horizontal wave numbers (see Eq. \ref{xh_WKBJ}), respectively) and mostly horizontal, i.e. $\xi_{r;n,l}\!<\!<\!\xi_{h;n,l}$ because of their weak compressibility. The inertia thus simplifies onto:
\begin{equation}
    I_{n,l} \approx l(l+1)\int_{r_i}^{r_o} |\xi_{h;n,l}|^2 \rho r^2 \,{\rm d}r, 
    \label{Inertia}
\end{equation}
where $r_i$ and $r_o$ are the inner and outer turning point, respectively \citep{Shibahashi1979}. We define $\zeta$ the ratio of the mode inertia within the cavity where it propagates as a gravity mode with $I$.
Following \cite{HekkerJCD2017} and \cite{Mathisetal2021}, the horizontal component of the displacement is expressed in this regime by its JWKB \citep[Jeffreys, Wentzel, Krammer, Brillouin;][]{FromanFroman2005} expression:  
\begin{equation}
    \xi_{h;n,l} \sim \rho^{-1/2} r^{-3/2} N^{1/2} \sin\left[\Phi_{n,l}\left(r\right)\right] \label{xh_WKBJ}
\end{equation}
with the phase $\displaystyle\Phi_{n,l}\left(r\right) =\int_{r_i}^r k_{r;n,l}(r') \,\mathrm{d}r' - \frac{\pi}{4}$ where $k_{r;n,l}=N/\omega_{0;n,l}\times k_{h;l}$ with $k_{h;l}\equiv\sqrt{l\left(l+1\right)}/r$.
In the case of magnetic topologies that vary over scales $L_{\rm B}$ larger than the radial wavelength of the mode, we demonstrate in Appendix \ref{Appen:Lorentz} that ${\mathcal L}_{\rm L}$ simplifies onto 
\begin{equation}
{\mathcal L}_{\rm L}\left[{\boldsymbol\xi}\right]
    =  \frac{1}{\rho\mu_0}\,B_{r}^2\left(r,\theta,\varphi\right)\,k_{r;n,l}^{2}\left(r\right)\,{\xi}_{h;n,l}\left(r\right){\boldsymbol S}_{l}^{m}\left(\theta,\varphi\right).
\end{equation}   
Following the rapid phase approximation for radial rapidly-oscillating integrals \citep{Mathisetal2021}, we obtain the general expression for $M_{m,m'}$:
\begin{eqnarray}
\lefteqn{M_{m,m'}=}\nonumber\\
&&\frac{1}{2\mu_0\omega_{0;m,l}^3}{\mathcal I}\int_{r_i}^{r_o}K\left(r\right)\int_{\Omega=4\pi}B_{r}^2\left({\boldsymbol S}_{l}^{m}\left(\theta,\varphi\right)\right)^{*}\cdot{\boldsymbol S}_{l}^{m'}\left(\theta,\varphi\right){\rm d}\Omega,\nonumber\\
\label{eq:matrixangular}
\end{eqnarray}
where ${\rm d}\Omega=\sin\theta{\rm d}\theta{\rm d}\varphi$, $\Omega$ being the solid angle, and
\begin{equation}
    K\left(r\right) = \frac{\displaystyle{\frac{1}{\rho} \left(\frac{N}{r}\right)^3}}{\displaystyle{ \int_{r_i}^{r_o} \left(\frac{N}{r}\right)^3 \frac{\hbox{d}r}{\rho} }}\quad\hbox{and}\quad\mathcal{I} =\frac{\displaystyle{\int_{r_i}^{r_o} \left(\frac{N}{r}\right)^3 \frac{\hbox{d}r}{\rho}}}{\displaystyle{\int_{r_i}^{r_o} \left(\frac{N}{r}\right) \,\hbox{d}r}}.
    \label{eq:kernel}
\end{equation}
The function $K$ peaks in a narrow range at the hydrogen burning shell because of the cubic dependency of $N/r$.

Finally, we use the Racah-Wigner algebra detailed in Appendix \ref{Appen:RW}, which is intensively used in Quantum mechanics, to reduce analytically the horizontal integral $\int_{\Omega=4\pi}B_{r}^2\left({\boldsymbol S}_{l}^{m}\left(\theta,\varphi\right)\right)^{*}\cdot{\boldsymbol S}_{l}^{m'}\left(\theta,\varphi\right){\rm d}\Omega$. Using Eqs. (\ref{conj}), (\ref{eq:SVHS}), (\ref{eq:6j}), and (\ref{eq:3j}), we finally obtain:
\begin{eqnarray}
\lefteqn{M_{m,m'}=\frac{1}{2\mu_0\omega_{0;m,l}^3}{\mathcal I}\sum_{l_1=1}^{L_B}\sum_{m_1=-l_1}^{l_1}\sum_{l_2=1}^{L_B}\sum_{m_2=-l_2}^{l_2}\sum_{L=\vert l_1-l_2\vert}^{l_1+l_2}\alpha_{l,m,m'}^{L}{\mathcal J}_{l_1,m_1,l_2,m_2}^{L,m'-m}}\nonumber\\
&&\times\delta_{m'-m,-\left(m_1+m_2\right)}\int_{r_i}^{r_o}K\left(r\right)b_{m_1}^{l_1}\left(r\right)b_{m_2}^{l_2}\left(r\right){\rm d}r,
\end{eqnarray}
with
\begin{eqnarray}
\lefteqn{\alpha_{l,m,m'}^{L}=\frac{\left(-1\right)^m}{2l+1}\left\{l\left(l+1\right)^2{\mathcal K}_{l,-m,-1,l,m',-1}^{L,m'-m}+\left(l+1\right)l^2{\mathcal K}_{l,-m,1,l,m',1}^{L,m'-m}\right.}\nonumber\\
&&+{\left.\left[l\left(l+1\right)\right]^{3/2}\left({\mathcal K}_{l,-m,-1,l,m',1}^{L,m'-m}+{\mathcal K}_{l,-m,1,l,m',-1}^{L,m'-m}\right)\right\}},
\end{eqnarray}
while the ${\mathcal K}$ and ${\mathcal J}$ coupling coefficients, which involve the 3j and 6j coefficients, have been defined in Eqs. (\ref{eq:6j}) \& (\ref{eq:3j}), respectively; $\delta$ is the usual Kronecker symbol. This will allow us in the next section to provide a precise physical diagnosis on the geometrical couplings between the (squared radial component of the) magnetic field and dipolar mixed modes.

Solving Eq. (\ref{eq:system}) leads to find $\left(2l+1\right)$ frequencies in the rotating frame $\omega_m=\omega_{0;n,l}+\omega_{1;m}$ ($m\in\left\{-l,\cdot\cdot\cdot,l\right\}$). Next, we take the Doppler-shift into account to get the frequencies $\omega_{m,m'} = \omega_m - m'\langle\Omega\rangle_{\rm g}$ in the inertial frame and finally define the frequency shifts
\begin{equation}
\delta\!\omega_{m,m'}=\omega_{m,m'}-\omega_{0;n,l}=\omega_{1;m}- m'\langle\Omega\rangle_{\rm g}.
\end{equation}

\subsection{The case of dipolar mixed modes}
\label{sec:Dipolar}

We focus on dipolar ($l=1$) mixed modes, which are those mainly observed in asteroseismology. In the more general case, its multiplet has nine components. Following \cite{Lietal2022}, we define the average of the frequency shifts of this multiplet as
\begin{equation}
    \delta\omega_{\rm B} = \frac{1}{9} \sum_{m,m'} \delta\omega_{m,m'}=\frac{1}{9}\sum_{m,m'} \omega_{1;m} - m'\langle\Omega\rangle_{\rm g}.
\end{equation}

In this case, the diagonal ${\boldsymbol R}$ matrix is given by
\begin{equation}
    R_{m,m} =  m \omega_{\rm R}\quad\hbox{with}\quad\omega_\mathrm{R} = \left(1-\frac{\zeta}{2}\right)\langle\Omega\rangle_{\rm g} - (1-\zeta)\langle\Omega\rangle_{\rm p},
\label{eq:rotation}
\end{equation}
which corresponds to the classical, symmetric when computed at the first-order, rotational splitting. Using the property that the sum of the eigenvalues of a matrix is its trace and that ${\rm Tr}\,{\boldsymbol R}=0$, we find using the results of \S \ref{sec:General} and Appendix \ref{Sec:AppendixDipolar} that
\begin{equation}
    \delta\omega_{\rm B} =\frac{1}{3}\zeta\,{\rm Tr}\,{\boldsymbol M}=\zeta\omega_{\rm B},
\end{equation}
with
\begin{equation}
\omega_{\rm B} = \frac{\mathcal{I}}{\mu_0\omega^{3}}\int_{r_i}^{r_o}K\left(r\right) \overline{B_r^2}{\rm d}r, \label{eq:omegaB}
\end{equation}
where we compute analytically the horizontal average over the sphere of the squared radial component of the  magnetic field using Eq. (\ref{eq:3j})
\begin{eqnarray}
    \lefteqn{\overline{B_r^2} = \frac{1}{4\pi}\int_{\Omega=4\pi}B_r^2{\rm d}\Omega}\nonumber\\
    &=&\frac{1}{4\pi}\sum_{l_1=1}^{L_B}\sum_{l_2=1}^{L_B}\sum_{m_1=-l_1}^{l_1}b_{m_1}^{l_1}\left(r\right)b_{-m_{1}}^{l_2}\left(r\right){\mathcal C}_{l_1,l_2,m_1}^{0},
\end{eqnarray}
with
\begin{equation}
{\mathcal C}_{l_1,l_2,m_1}^{L}=\sqrt{\left(2 l_1+1\right)\left(2 l_2 +1\right)}\left(\begin{array}{ccc}
            l_1 & l_2 & L\\
            0 & 0 & 0 \end{array}\right)\left(\begin{array}{ccc}
            l_1 & l_2 & L \\
            m_1 & -m_1 & 0 \end{array}\right),
\end{equation}
where $\left(\cdot\cdot\cdot\right)$ is a 3j coefficient.

In the case where the rotation terms dominate the magnetic ones, the nine components multiplet reduces to a triplet as in the case of the observed stars KIC 8684542, KIC 11515377 and KIC 7518143 studied in \cite{Lietal2022}. As demonstrated in \cite{Bugnetetal2021} and \cite{Lietal2022}, the presence of the magnetic field induces an asymmetry of the triplet 
\begin{equation}
\delta_{\rm asym}=\omega_{m=-1}+\omega_{m=1}-2\omega_{m=0}=2 \zeta \left(M_{11}-M_{00}\right)=3 \zeta a \omega_{\rm B},
\end{equation}
where we introduce the asymmetry parameter $a$
\begin{eqnarray}
\lefteqn{a = \frac{ \displaystyle \int_{r_i}^{r_o} K(r) \int_{\Omega=4\pi} B_r^2 P_2(\cos\theta)\,{\rm d}\Omega \,{\rm d}r} {\displaystyle \int_{r_i}^{r_o} K(r) \int_{\Omega=4\pi} B_r^2 \,{\rm d}\Omega \,{\rm d}r }}\nonumber\\
&&=\frac{\displaystyle{\sum_{l_1=1}^{L_B}\sum_{l_2=1}^{L_B}\sum_{m_1=-l_1}^{l_1}{\mathcal C}_{l_1,l_2,m_1}^{2}\int_{r_i}^{r_o}K\left(r\right)b_{m_1}^{l_1}\left(r\right)b_{-m_{1}}^{l_2}\left(r\right){\rm d}r}}{\displaystyle{\sum_{l_1=1}^{L_B}\sum_{l_2=1}^{L_B}\sum_{m_1=-l_1}^{l_1}{\mathcal C}_{l_1,l_2,m_1}^{0}\int_{r_i}^{r_o}K\left(r\right)b_{m_1}^{l_1}\left(r\right)b_{-m_{1}}^{l_2}\left(r\right){\rm d}r}},
\label{eq:a}
\end{eqnarray}
which we compute analytically using the Racah-Wigner algebra; $P_{2}\left(\cos\theta\right)$ is the Legendre polynomial of order $l=2$. It allows us in \S \ref{sec:topology} to identify the cases where the measurement of $\delta_{\bf asym}$ allows us to unravel and characterise the properties of the topology of the field and those where it is blind or where it only provides partial information. Since $-1/2\le P_2\left(\cos\theta\right)\le 1$, we also have $-1/2<a<1$ where the minimum and maximum $a$ values correspond to the cases where $B_r^2$ is totally concentrated around the pole or along the equator, respectively.

\section{An exploration of possible magnetic topologies}
\label{sec:topology}

\subsection{The classical oblique dipolar fossil field}
\label{subsec:obliquedipole}

The classical model for a fossil field in a stellar radiation zone is the oblique rotator model \citep[e.g.][]{Moss1980,Mesteletal1981} in which a dipolar field is inclined with respect to the rotation axis with an angle $\beta$ between the direction of the magnetic dipolar moment and the rotation axis. Following \cite{Pratetal2020}, its radial component is given by:
\begin{equation}
B_r\left(r,\theta,\varphi\right)=B_{0}\,b_{r}\left(r\right)\left(\cos\beta\cos\theta+\sin\beta\sin\theta\cos\varphi\right),
\end{equation}
where $B_{0}$ is the field amplitude and $b_{r}\left(r\right)$ its radial distribution.
When applying the general result given in Eq. (\ref{eq:a}), we obtain:
\begin{equation}
a=\frac{1}{5}\left(2\cos^2\beta-\sin^2\beta\right)=\frac{2}{5}P_{2}\left(\cos\beta\right)=\frac{2}{5}d_{0,0}^{2}\left(\beta\right),
\label{eq:aobliquedipole}
\end{equation}
where $d_{m',m}^{l}\left(\cos\beta\right)$ is the rotation matrix used in quantum mechanics \citep{Varshalovich1988} to project a spherical harmonics $Y_{l}^{m}\left(\theta,\varphi\right)$ on spherical harmonics $Y_{l}^{m'}\left(\theta',\varphi'\right)$ when the vertical axis of the reference frame, in which the spherical coordinates $\left(\theta,\varphi\right)$ are defined, is rotated by an angle $\beta$. This leads to a new reference frame with its spherical coordinates $\left(\theta',\varphi'\right)$. If we consider $B_r^2$ in the reference frame of the magnetic dipole, it projects on axisymmetric spherical harmonics $Y_{0}^{0}\left(\theta_{\rm B}\right)$ and $Y_{2}^{0}\left(\theta_{\rm B}\right)$, where $\theta_{\rm B}$ is the co-latitude in this frame, because $B_r^2\propto\cos^2\theta_{\rm B}$. When rotating back to the reference frame with the vertical axis aligned along the rotation axis, we are thus doing a rotation with the angle $\beta$. This explains why $a\propto d_{0,0}^{2}\left(\beta\right)$ since we project $B_{r}^{2}$ on the spherical harmonics $Y_{2}^{0}\left(\theta\right)\propto P_2\left(\cos\theta\right)$. When we plot $a$ as a function of $\beta$ we recover the results reported in \cite{Lietal2022} that $a=2/5$ when $\beta=0^{\circ}$ and $a=-1/5$ when $\beta=90^{\circ}$. For a given measured asymmetry $a$, we can invert the previous equation that leads to
\begin{equation}
\beta= \frac{1}{2}\arccos\left(\frac{10a-1}{3}\right).
\label{eq:asymangle}
\end{equation}
We thus find the critical inclination $\beta=54.736^{\circ}$ for which the sign of $a$ changes.

\begin{figure}[t!]
   \centering
   \includegraphics[width=9.25cm]{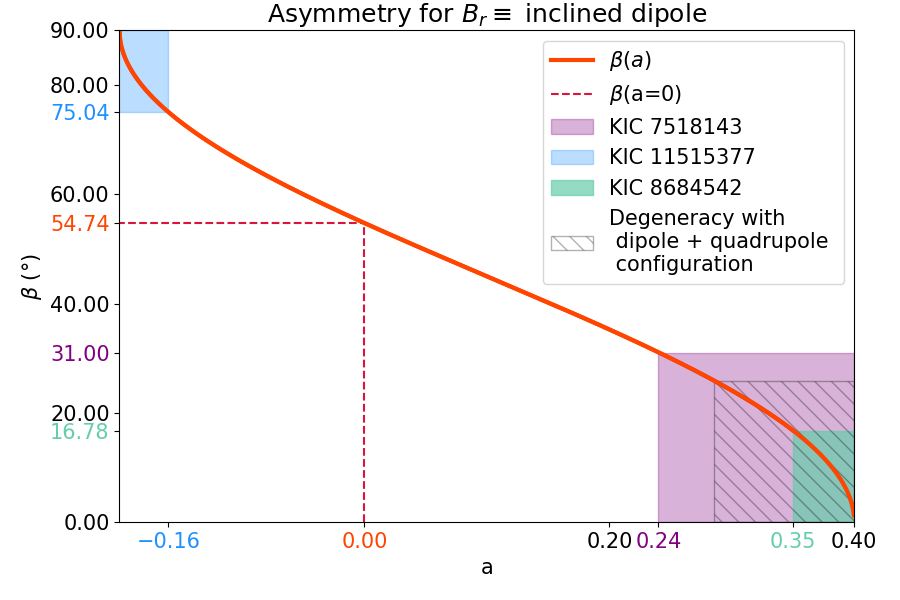}
      \caption{Inclination angle of the dipole field as function of the measured asymmetry. Possible inclination angles for KIC 8684542, KIC 11515377, and KIC 7518143 are represented.}
         \label{Fig:ObliqueRotator}
   \end{figure}

We can apply this result to the three stars observed by \cite{Lietal2022} KIC 8684542, KIC 11515377 and KIC 7518143. For KIC 8684542, we get $0^{\circ}<\beta<16.78^{\circ}$ since $a=0.47\pm 0.12$. For KIC 11515377, we obtain $75.04^{\circ}<\beta<90^{\circ}$ since $a=-0.24\,+0.08/-0.234$. Finally, for KIC 7518143, we get that $\beta<31^{\circ}$ since $a>0.24$. These results are represented on Fig.~\ref{Fig:ObliqueRotator}. These applications demonstrate how helpful and powerful is the Racah-Wigner algebra used here to compute analytically $a$ as a function of the considered inclined dipolar magnetic topology and to derive its obliquity. However, one should remember that this is derived assuming we know the type of magnetic topology we are looking for.

\subsection{Combining dipolar and quadrupolar fields}
\begin{figure}[th!]
   \centering
\includegraphics[width=8.75cm]{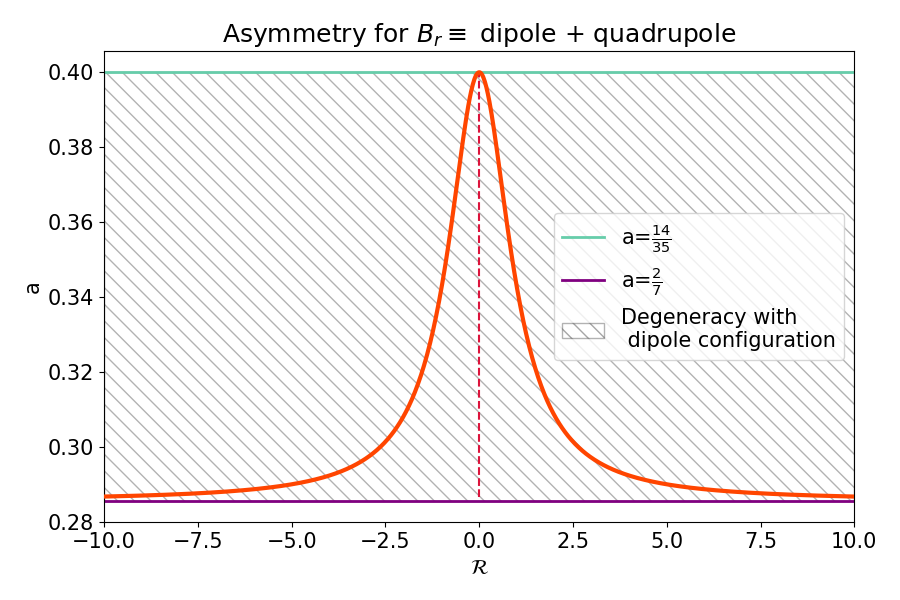}
      \caption{Asymmetry parameter as function of the ratio between the dipolar and the quadrupolar components in the case of a dipolar + quadrupolar field.}
         \label{Fig_cartoon}
\end{figure}

To investigate the possible degeneracy of the asymmetry of mixed modes as a diagnosis of the magnetic topology, we consider an axisymmetric topology composed by a combination of a dipolar and a quadrupolar components \citep[e.g.][]{Moss1974,Moss1985}. We choose a topology defined as
\begin{equation}
B_{r}\left(r,\theta\right)=B_{0}\,b\left(r\right)\left[Y_{1}^{0}\left(\theta\right)+{\mathcal R} Y_{2}^{0}\left(\theta\right)\right],
\end{equation}
where ${\mathcal R}$ is the ratio between the dipolar and the quadrupolar components. We get
\begin{equation}
a=\frac{2}{7}+\frac{4}{35}\frac{1}{1+{\mathcal R}^2},
\label{eq:adipquad}
\end{equation}
which we plot hereafter. When comparing the obtained values with those in Fig. \ref{Fig:ObliqueRotator}, we see from Fig.~\ref{Fig_cartoon} that it is not objectively possible to disentangle between the combination proposed here with an oblique dipole with an angle $\beta<\approx 25.88^{\circ}$ computed using Eq. (\ref{eq:asymangle}) with the minimum asymptotic value obtained in Eq. (\ref{eq:adipquad}), i.e. $a=2/7$.   

\subsection{The case of a general field}

\begin{figure}[h!]
   \centering
   \includegraphics[width=8.75cm]{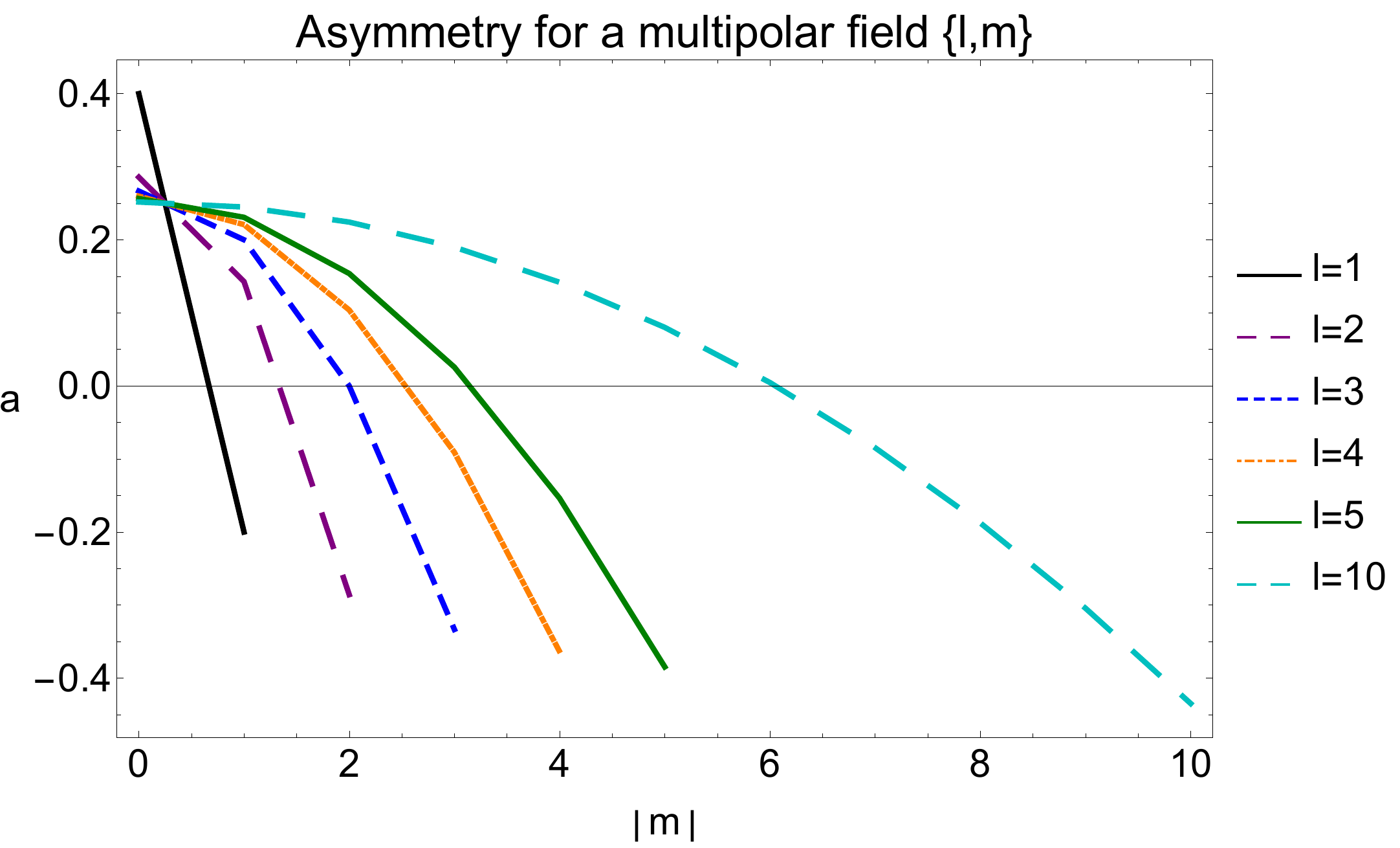}
      \caption{Asymmetry parameter for a given magnetic field with a latitudinal degree $1 \le l \le 10$ and an azimuthal order $-l \le m \le l$.}
         \label{Fig:asymMultipole}
   \end{figure}

In the case of the oblique dipolar fossil field, we have been able to demonstrate explicitly in Sec. \ref{subsec:obliquedipole} how $a$ is providing a very precise constrain on the inclination angle $\beta$ of the field. In the case where we have combined axisymmetric dipolar and quadripolar fields, we have shown how $a$ becomes partially blind and a degenerated diagnosis parameter when we don't know a priori the type of configurations we are looking for. It is thus now interesting to consider a modal magnetic field such as $B_r\left(r,\theta,\varphi\right)=B_{0}\,b\left(r\right)\left[Y_{l}^{m}\left(\theta,\varphi\right)+Y_{l}^{-m}\left(\theta,\varphi\right)\right]$ to unravel the potential information provided by $a$ and its sign. In Figure \ref{Fig:asymMultipole}, we represent the variation of $a$ as a function of $\vert m\vert$, since $a(-m)=a(m)$, for a fixed $l$. We verify that $-1/2 \le a \le 1$. We see that for each $l$, there is a critical $m$ above which the sign of $a$ changes and becomes negative. This is an interesting result, as it shows that in the case of multipolar magnetic field where a $\left\{l,m\right\}$ mode dominates the magnetic energy spectrum, the sign of $a$ provides us an information on how close the field is to a sectoral (i.e. $l=m$) configuration, and thus its degree of non-axisymmetry. Combined with results obtained thanks to MHD numerical simulations \citep[e.g.][]{BraithwaiteSpruit2004,BraithwaiteNordlund2006,Brun2007,Braithwaite2008,EmeriauBrun2017,Becerraetal2022} on fossil field formation and instabilities, the sign of the asymmetry parameter could therefore provide us precious information on the initial magnetic conditions and the stability of the observed fields.    

\section{Conclusions}
In this letter, we put to use the powerful Racah-Wigner algebra used in Quantum mechanics to explore the behavior of the asymmetry of the frequency splittings of dipolar mixed gravito-acoustic modes when they propagate in the magnetised radiative core of an evolved low- or intermediate-mass star. First, following \cite{Lietal2022}, we studied the classic case of a dipolar fossil field, which can have any inclination angle $\beta$ with respect to the rotation axis. The Racah-Wigner algebra allows us to provide an exact analytical expression of the asymmetry as a function of this angle (Eq. \ref{eq:aobliquedipole}). This allows us, thanks to Eq. \ref{eq:asymangle} to constrain its value using the observed asymmetry for the observed stars KIC 8684542, KIC 11515377 and KIC 7518143 (in Fig. \ref{Fig:ObliqueRotator}). Next, we explore the behavior of a more complex, but potentially less stable fossil field configuration\footnote{\cite{DuezMathis2010} have demonstrated that the most stable lowest-energy mixed (poloidal + toroidal) equilibrium configuration is the dipolar one, while higher degree equilibrium configurations have higher energy and are thus less stable.}, which has an axisymmetric quadrupolar component in addition to an axisymmetric dipole. We show that the signature of such field lead to similar values of the asymmetry that the one obtained with a sole dipolar field with an angle $\beta<\approx 25.88^{\circ}$. Finally, we explore the more general case of a multipolar field for which a $\left\{l,m\right\}$ mode dominates the magnetic energy spectrum. We show that the asymmetry parameter becomes negative when $m\rightarrow l$. In other word, a negative asymmetry is the sign of a strong non-axisymmetric field. This study therefore demonstrates the difficulty to constrain the possible geometrical configurations of deep fossil fields, and the importance to develop simultaneously asteroseismic diagnosis and MHD simulations of their formation and instabilities to unravel their main properties.

\begin{acknowledgements}
     The authors are grateful to the referee for her/his detailed and constructive report, which has allowed us to improve our article. S. M. acknowledges support from the CNES GOLF-SOHO and PLATO grants at CEA/DAp and PNPS (CNRS/INSU). We thank R. A. Garcia for fruitful discussions and suggestions.
\end{acknowledgements}

\bibliographystyle{aa}
\bibliography{Mathis2022}

\begin{appendix}

\section{Asymptotic expression of the Lorentz force}
\label{Appen:Lorentz}
To compute the impact of deep magnetic fields on mixed modes, we can focus on the regime of asymptotic g modes, which are rapidly oscillating along the radial direction (we refer the reader to Sec. \ref{sec:General}).

The Lorentz force can be expressed in the following two ways:
\begin{equation}
{\boldsymbol F}_{\rm L}=\frac{1}{\mu_0}\left({\boldsymbol \nabla}\times{\boldsymbol B}\right)\times{\boldsymbol B}=\frac{1}{\mu_0}\left({\boldsymbol B}\cdot{\boldsymbol \nabla}\right){\boldsymbol B}-{\boldsymbol\nabla}\left(\frac{{\boldsymbol B}^2}{2\mu_0}\right),
\end{equation}
where ${\boldsymbol B}$ is the magnetic field. We recognise in the second expansion, the magnetic tension force and the gradient of the magnetic pressure. 

We introduce the linear expansion of ${\boldsymbol B}$:
\begin{equation}
{\boldsymbol B}={\boldsymbol B}_{0}+{\boldsymbol B}',
\end{equation}
where ${\boldsymbol B}_{0}$ and ${\boldsymbol B}'$ are the background magnetic field and the wave-induced magnetic field perturbation, respectively. The latter is computed using the linearised induction equation in its adiabatic limit:
\begin{eqnarray}
\lefteqn{\partial_{t}{\boldsymbol B}'={\boldsymbol\nabla}\times\left({\boldsymbol u}\times{\boldsymbol B}_{0}\right)}\nonumber\\
&=&\left({\boldsymbol B}_{0}\cdot{\boldsymbol\nabla}\right){\boldsymbol u}-\left({\boldsymbol u}\cdot{\boldsymbol\nabla}\right){\boldsymbol B}_{0}-{\boldsymbol B}_{0}\left({\boldsymbol\nabla}\cdot{\boldsymbol u}\right),
\label{eq:LinInduc}
\end{eqnarray}
where ${\boldsymbol u}=\partial_{t}{\boldsymbol\xi}$ is the velocity field of the oscillation mode with ${\boldsymbol\xi}$ its Lagrangian displacement; we omit in this appendix any $_{n,l}$ subscripts to lighten notations. Since mixed modes we are considering are behaving as asymptotic g-modes in radiative layers, they are nearly incompressible (i.e. ${\boldsymbol \nabla}\cdot{\boldsymbol u}\approx 0$), predominantly horizontal (i.e. $\xi_r\!<\!\!<\!\xi_h$), and rapidly oscillating in the radial direction (i.e. $k_r\!>\!\!>\!k_h$). If we focus on background magnetic topologies ($\boldsymbol B_{0}$), which have characteristic length of variation larger than the mode wavelength, Eq. (\ref{eq:LinInduc}) simplifies onto:
\begin{equation}
\partial_{t}{\boldsymbol B}'\approx\left({\boldsymbol B}_{0}\cdot{\boldsymbol\nabla}\right){\boldsymbol u}.
\end{equation}
It leads thanks to the JWKB approximation, which applies for  asymptotic g-modes, to:
\begin{equation}
{\boldsymbol B}'\approx B_r\left(r,\theta,\varphi\right)\frac{k_r}{\omega_0}{\boldsymbol u}_h,
\label{eq:mastersimplify}
\end{equation}
where ${\boldsymbol u}_h=\partial_t{\boldsymbol \xi}_h=i{\omega_0}{\boldsymbol \xi}_h$.

The linearised Lorentz force
\begin{equation}
{\boldsymbol F}_{\rm L}=\frac{1}{\mu_0}\left[\left({\boldsymbol B}_{0}\cdot{\boldsymbol \nabla}\right){\boldsymbol B}'+\left({\boldsymbol B}'\cdot{\boldsymbol \nabla}\right){\boldsymbol B}_0\right]-{\boldsymbol\nabla}\left(\frac{{\boldsymbol B}_0\cdot{\boldsymbol B}'}{\mu_0}\right)
\end{equation}
also simplifies into:
\begin{equation}
{\boldsymbol F}_{\rm L}\approx\frac{1}{\mu_0}\left({\boldsymbol B}_{0}\cdot{\boldsymbol \nabla}\right){\boldsymbol B}',
\end{equation}
where we have assumed that the fluctuations of the magnetic pressure are negligible when compared to those of the gaseous pressure. Using Eq. (\ref{eq:mastersimplify}) and the JWKB approximation, we finally obtain:
\begin{equation}
\frac{{\boldsymbol F}_{\rm L}}{\rho}=-\frac{B_r^2\left(r,\theta,\varphi\right)}{\rho\,\mu_0}k_{r}^2{\boldsymbol\xi}_{h}=-{\mathcal L}_{\rm L}\left[\vec\xi\right],
\label{eq:finallinlorentz}
\end{equation}
where we have introduced the radial Alfv\'en velocity $V_{r;{\rm A}}=B_r/\sqrt{\rho\,\mu_0}$ and the corresponding pulsation $\omega_{\rm A}=V_{r;{\rm A}} k_{r}$. The ${\mathcal L}_{\rm L}$ operator corresponds to the so-called direct modification of modes' eigenfrequencies and eigenfunctions by the linearized Lorentz force \citep[][]{GoughThompson1990}. It scales as
\begin{equation}
R_{\rm direct}=\frac{\vert\vert{\mathcal L}_{\rm L}\left[\boldsymbol\xi\right]\vert\vert}{\vert\vert\omega_{0}^2{\boldsymbol\xi}\vert\vert}=\frac{\vert\vert{\mathcal L}_{\rm L}\left[\boldsymbol\xi\right]\vert\vert}{\vert\vert{\mathcal L}_{0}\left[\boldsymbol\xi\right]\vert\vert}\approx\left(\frac{\omega_{\rm A}}{\omega_0}\right)^2
\end{equation}
when compared to the mode acceleration in the non-magnetic case.\\ 

However, indirect effects can also affect the modes. In the case of a non force-free field, as this is expected for fossil fields \citep[][]{Reisenegger2009,DuezMathis2010}, the hydrostatic structure of the star is distorted by the Lorentz force \citep[][]{DMTC2010,FullerMathis2023}. This modifies the cavity propagation of the modes and thus also contributes to the perturbation of their eigenfrequencies and eigenfunctions. We refer the reader to the seminal article by \cite{GoughThompson1990} for the complete mathematical modeling of this effect. We here focus on its order of magnitude. It scales as:
\begin{equation}
R_{\rm indirect}=\frac{\left\lvert\left\lvert\displaystyle{\frac{\left({\boldsymbol \nabla}\times{\boldsymbol B_{0}}\right)\times{\boldsymbol B_{0}}}{\rho\mu_{0}}}\right\rvert\right\rvert}{\vert\vert{\boldsymbol g}\vert\vert}\approx\frac{V_{\rm A}^2/r_{\rm c}}{\left(G M_{r}\left(r_c\right)\right)/r_c^2},
\end{equation}
where we introduce the Alfv\'en velocity, $V_{\rm A}=\vert\vert{\boldsymbol B}_{0}\vert\vert/\sqrt{\rho\mu_0}$, the Universal gravity constant, $G$, the mass contained in a sphere of radius $r$, $M_r$, and the radius of the radiative core, $r_c$, which we choose as the characteristic length of variation of the fossil field.

We evaluate the ratio between the direct and the indirect effects induced by the magnetic field in the case of dipolar mixed modes:
\begin{equation}
\frac{R_{\rm direct}}{R_{\rm indirect}}\approx 2 \left(\frac{N}{\omega_{0}}\right)^{2}\left(\frac{\omega_{\rm d}\left(r_c\right)}{\omega_{0}}\right)^{2},
\end{equation}
where $\omega_{\rm d}=\sqrt{\left(G M_r\left(r_c\right)\right)/r_c^3}$ is the dynamical frequency and we have assumed that $V_{r;{\rm A}}$ and $V_{\rm A}$ have the same order of magnitude. 

Computing models of typical red giants using the MESA stellar structure and evolution code \citep{Paxtonetal2011}, we can evaluate this ratio. For a $1.5$M$_{\odot}$ star with a metallicity $Z=0.02$, we have $\omega_0\approx 2 \pi \nu_{\rm max}$ with $\nu_{\rm max}\approx 150\mu{\rm Hz}$, $N_{\rm max}\approx 10^{4}\mu{\rm Hz}$, $r_c\approx0.75$R$_{\odot}$, $M_r\left(r_c\right)\approx 0.3$M$_{\odot}$, where R$_{\odot}$ and M$_{\odot}$ are the Solar radius and mass, respectively, that leads to $R_{\rm direct}/R_{\rm indirect}\approx 2803\!>\!\!>\!1$. This justifies why the indirect effect has been neglected in studies devoted to red giant stars. \\

In addition, when considering the full mathematical expression of the perturbation of the wave operator induced by indirect effects \citep[we refer the reader to Eq. 3.24 in][]{GoughThompson1990}, we can identify that its dominant terms, which are those with the highest derivatives of the eigenfunction because they are rapidly oscillating along the radial direction in the asymptotic regime, scale with the compressibility of the mode, which is weak for high-order g-modes. The second term induced by indirect effects scales as $\left(\rho_{\rm B}/\rho\right)\omega_{0}^{2}$, where $\rho_B$ is the perturbation of the hydrostatic background density induced by the Lorentz force \citep[see Eq. 2.27 in][]{GoughThompson1990}. As this can be seen for instance in \cite{DMTC2010}, the ratio $\rho_{\rm B}/\rho$ is very small in the deep central layers of stars.\\

Therefore, we can focus in our work on the direct effects and neglect the indirect terms.

\section{Spherical harmonics and Racah-Wigner algebra}
\label{Appen:RW}

\subsection{Scalar quantities}

\subsubsection{Definition and properties}
The  spherical harmonics of degree $l$ and order $m$ are defined by:
\begin{equation}
Y_{l}^{m}(\theta,\varphi)=\mathcal{N}_{l}^{m}P_{l}^{|m|}\left(\cos\theta\right)e^{im\varphi} ,
\label{eq:defSH}
\end{equation}
where $P_{l}^{|m|}\left(\cos\theta\right)$ is the associated Legendre polynomial, and the normalization coefficient is 
\begin{equation}
\mathcal{N}_{l}^{m}=(-1)^{\frac{\left(m+|m|\right)}{2}}\left[\frac{2l+1}{4\pi}\frac{(l-|m|)!}{(l+|m|)!}\right]^{\frac{1}{2}}.
\label{norm}
\end{equation}
They obey the orthogonality relation:
\begin{equation}
\int_{\Omega}\left(Y_{l_{1}}^{m_{1}}\left(\theta,\varphi\right)\right)^{*}Y_{l_{2}}^{m_{2}}\left(\theta,\varphi\right){\rm d}\Omega=\delta_{l_{1},l_{2}}\delta_{m_{1},m_{2}},
\label{ortho}
\end{equation}
where ${\rm d}\Omega=\sin\theta \, {\rm d}\theta \, {\rm d}\varphi$ and the complex conjugate spherical harmonics is given by:
\begin{equation}\left(Y_{l}^{m}\left(\theta,\varphi\right)\right)^{*}=\left(-1\right)^{m}Y_{l}^{-m}\left(\theta,\varphi\right).
\label{conj}
\end{equation}
Using these properties, every function $f(\theta,\varphi)$ can be expanded as:
\begin{equation}
f\left(\theta,\varphi\right)=\sum_{l=0}^{\infty}\sum_{m=-l}^{l}f_{m}^{l}Y_{l}^{m}\left(\theta,\varphi\right),
\label{exp}
\end{equation}
where
\begin{equation}
f_{m}^{l}=\int_{\Omega}f(\theta,\varphi)\left(Y_{l}^{m}\left(\theta,\varphi\right)\right)^{*}{\rm d}\Omega.
\end{equation}

\subsubsection{Products of spherical harmonics}
Using the normalization and the orthogonality of spherical harmonics (Eqs. \ref{ortho} \& \ref{exp}) and their complex conjugate (Eq. \ref{conj}), we can write: 
\begin{eqnarray}
\lefteqn{Y_{l_{1}}^{m_{1}}\left(\theta,\varphi\right)Y_{l_{2}}^{m_{2}}\left(\theta,\varphi\right)}\nonumber\\
&&=(-1)^{\left(m_{1}+m_{2}\right)}\sum_{l=|l_{1}-l_{2}|}^{l_{1}+l_{2}}\mathcal{J}_{l_{1},m_{1},l_{2},m_{2}}^{l,-\left(m_{1}+m_{2}\right)}Y_{l}^{m_{1}+m_{2}}\left(\theta,\varphi\right),
\end{eqnarray}
where we define the integral $\mathcal{J}_{l_{1},l_{2},l}^{m_{1},m_{2},m}$ following \cite{Varshalovich1988}:
\begin{eqnarray}
\lefteqn{\mathcal{J}_{l_{1},m_{1},l_{2},m_{2}}^{l,m}=
\int_{\Omega}Y_{l_{1}}^{m_{1}}\left(\theta,\varphi\right)Y_{l_{2}}^{m_{2}}\left(\theta,\varphi\right)Y_{l}^{m}\left(\theta,\varphi\right)d\Omega=}\nonumber\\
&=&\sqrt{\frac{(2l_{1}+1)(2l_{2}+1)(2l+1)}{4\pi}}\left(\begin{array}{ccc}
l_{1} & l_{2} & l \\
m_{1} & m_{2} & m
\end{array}\right)\nonumber\\
& &\times\left(\begin{array}{ccc}
l_{1} & l_{2} & l\\
0 & 0 & 0
\end{array}\right)
\label{eq:3j}
\end{eqnarray}
with the 3j-Wigner coefficients, which are related to the classical Clebsch-Gordan coefficients $C_{l_1,m_1,l_2,m_2}^{l,m}$ by: 
\begin{equation}
\left(\begin{array}{ccc}
l_{1} & l_{2} & l \\
m_{1} & m_{2} & m 
\end{array}\right)=\frac{(-1)^{l_{1}-l_{2}-m}}{\sqrt{2l+1}}C_{l_{1},m_{1},l_{2},m_{2}}^{l,-m}.
\end{equation}

\subsection{Vector fields}

\subsubsection{Spheroidal and toroidal basis}
Following \cite{Rieutord1987}, we expand any vector field ${\boldsymbol X}(r,\theta,\varphi)$ in vectorial spherical harmonics as
\begin{eqnarray}
\lefteqn{{\boldsymbol X}(r,\theta,\varphi)=}\nonumber\\
&&\sum_{l=0}^{\infty}\sum_{m=-l}^{l}\left\{u_{m}^{l}(r){\boldsymbol R}_{l}^{m}(\theta,\varphi)+v_{m}^{l}(r){\boldsymbol S}_{l}^{m}(\theta,\varphi)+w_{m}^{l}(r){\boldsymbol T}_{l}^{m}(\theta,\varphi)\right\},\nonumber\\
\label{vec}
\end{eqnarray}
where we introduce:
\begin{eqnarray}
{\boldsymbol R}_{l}^{m}(\theta,\varphi)&=&Y_{l}^{m}(\theta,\varphi){\boldsymbol{\widehat e}}_{r},\\
{\boldsymbol S}_{l}^{m}(\theta,\varphi)&=&{\boldsymbol\nabla}_{\rm H}Y_{l}^{m}(\theta,\varphi),\\
{\boldsymbol T}_{l}^{m}(\theta,\varphi)&=&{\boldsymbol\nabla}_{\rm H}\wedge{\boldsymbol R}_{l}^{m}(\theta,\varphi)
\end{eqnarray}
with the horizontal gradient ${\boldsymbol\nabla}_{\rm H}={\boldsymbol{\widehat e}}_{\theta}\partial_{\theta}+{\boldsymbol{\widehat e}}_{\varphi}\frac{1}{\sin\theta}\partial_{\varphi}$.
These vector functions obey the following orthogonality relations:
\begin{equation}
\int_{\Omega}\boldsymbol R_{l_{1}}^{m_{1}}\cdot\boldsymbol S_{l_{2}}^{m_{2}}d\Omega=\int_{\Omega}\boldsymbol R_{l_{1}}^{m_{1}}\cdot\boldsymbol T_{l_{2}}^{m_{2}}d\Omega=\int_{\Omega}\boldsymbol S_{l_{1}}^{m_{1}}\cdot\boldsymbol T_{l_{2}}^{m_{2}}d\Omega=0,
\end{equation}
\begin{equation}
\int_{\Omega}\boldsymbol R_{l_{1}}^{m_{1}}\cdot\left(\boldsymbol R_{l_{2}}^{m_{2}}\right)^{*}{\rm d}\Omega=\delta_{l_{1},l_{2}}\delta_{m_{1},m_{2}}
\end{equation}
\begin{equation}
\int_{\Omega}\boldsymbol S_{l_{1}}^{m_{1}}\cdot\left(\boldsymbol S_{l_{2}}^{m_{2}}\right)^{*}d\Omega=\int_{\Omega}\boldsymbol T_{l_{1}}^{m_{1}}\cdot\left(\boldsymbol T_{l_{2}}^{m_{2}}\right)^{*}d\Omega=l_{1}(l_{1}+1)\delta_{l_{1},l_{2}}\delta_{m_{1},m_{2}}.
\end{equation}
We note that $\boldsymbol R_{l}^{m}$ and $\boldsymbol S_{l}^{m}$ represent the poloidal basis, and $\boldsymbol T_{l}^{m}$ the toroidal basis of $\boldsymbol X$.

\subsubsection{Spin vector harmonics}

\textit{Definition and properties:}\\

Unfortunately, if the $\left({\boldsymbol R},{\boldsymbol S},{\boldsymbol T}\right)$ basis is very well adapted to compute linear spectral projections, for instance of linear differential operators \citep{Rieutord1987}, it fails for computing the spectral projections of scalar products and of vectorial products. Since we have here to compute the projection of ${\boldsymbol S}_{l}^{m}\left(\theta,\varphi\right)\cdot\left({\boldsymbol S}_{l}^{m'}\left(\theta,\varphi\right)\right)^{*}$ on spherical harmonics in Eq. (\ref{eq:matrixangular}), we thus follow \cite{Varshalovich1988,Strugareketal2013} and we introduce the spin-1 vector harmonics (SVH). They are defined by:
\begin{eqnarray}
\lefteqn{\boldsymbol{Y}_{l,l+\nu}^{m}(\theta,\varphi)=}\nonumber\\
&&(-1)^{l-m}\!\sqrt{2l+1}\!
\sum_{\mu=-1}^{1}
\left(
\begin{array}{ccc}
l & l+\nu & 1 \\ 
m & -q & -\mu
\end{array}
\right) Y_{l+\nu}^{q}\left(\theta,\varphi\right){\boldsymbol{\widehat e}}_{\mu},\,\label{eqn:SVHdef}
\end{eqnarray}
with $q=m-\mu$ and the directional unit vectors:
\begin{eqnarray}
&\left\{ 
\begin{array}{l@{\quad}l}
{\boldsymbol{\widehat e}}_{-1}=\frac{1}{\sqrt{2}}\left({\boldsymbol{\widehat e}}_{x}-i\,{\boldsymbol{\widehat e}}_{y}\right) &  \\ 
{\boldsymbol{\widehat e}}_{0}={\boldsymbol{\widehat e}}_{z} &  \\ 
{\boldsymbol{\widehat e}}_{+1}=-\frac{1}{\sqrt{2}}\left({\boldsymbol{\widehat e}}_{x}+i\,{\boldsymbol{\widehat e}}_{y}\right). & 
\end{array}
\right.\label{eqn:SVHunivec}
\end{eqnarray}
They form an orthonormal basis over the space of all vector functions that satisfy the vector Laplace equation in spherical coordinates. Their orthonormality condition follows from the properties of the scalar spherical harmonics:
\begin{equation}
\int_{\Omega}\boldsymbol{Y}_{l_{1},l_{1}+\nu_{1}}^{m_{1}}\left(\theta,\varphi
\right)\cdot\left(\boldsymbol{Y}_{l_{2}+\nu_{2}}^{m_{2}}\left(\theta,\varphi\right)
\right)^{*}d\Omega=\delta_{l_{1},l_{2}}\delta_{m_{1},m_{2}}\delta_{\nu_{1},
\nu_{2}}.
\end{equation}
As in the $\left({\boldsymbol R},{\boldsymbol S},{\boldsymbol T}\right)$ basis, any vector field ${\boldsymbol X}\left(r,\theta,\varphi\right)$ can be expanded on the SVH basis as:
\begin{equation}
\boldsymbol
X\left(r,\theta,\varphi\right)=\sum_{l=0}^{\infty}\sum_{m=-l}^{l}\sum_{
\nu=-1}^{1}X_{l,l+\nu}^{m}(r)\boldsymbol Y_{l,l+\nu}^{m}\left(\theta,\varphi\right),
\end{equation}
with the radial-dependent coefficients
\begin{equation}
X_{l,l+\nu}^{m}(r)=\int_{\Omega}\left(\boldsymbol
Y_{l,l+\nu}^{m}\left(\theta,\varphi\right)\right)^{*}\cdot\boldsymbol
X\left(r,\theta,\varphi\right)d\Omega.
\end{equation}

\noindent\textit{Scalar product:}\\

Following \cite{Varshalovich1988}, we can compute the projection of a scalar product of two SVHs on spherical harmonics as follows:
\begin{equation}
\boldsymbol Y_{l_{1},l_{1}+\nu_{1}}^{m_{1}}\left(\theta,\varphi\right)\cdot\boldsymbol
Y_{l_{2},l_{2}+\nu_{2}}^{m_{2}}\left(\theta,\varphi\right)=
\sum_{l}\mathcal{K}
_{l_{1},m_{1},\nu_{1},l_{2},m_{2},\nu_{2}}^{l,m_{1}+m_{2}}Y_{l}^{m_{1}+m_{2}}\left(\theta,\varphi\right),
\label{eq:SVH_SP}
\end{equation}
with 
\begin{eqnarray}
\lefteqn{\mathcal{K}_{l_{1},m_{1},\nu_{1},l_{2},m_{2},
\nu_{2}}^{l,m_{1}+m_{2}}=(-1)^{\left[l_{1}-(l_{2}+\nu_{2})+l+m\right]}}\nonumber\\ 
& &\sqrt{\frac{\left(2l_1+1\right)\left(2l_2+1\right)\left(2l_1+2\nu_1+1\right)\left(2l_2+2\nu_2+1\right)\left(2l+1\right)}{4\pi}}\nonumber\\
& &\left\{
\begin{array}{ccc}
l_{1}+\nu_{1} & l_{2}+\nu_{2} & l \\ 
l_{2} & l_{1} & 1
\end{array}
\right\} \left(
\begin{array}{ccc}
l_{1} & l_{2} & l \\ 
m_{1} & m_{2} & -\left(m_{1}+m_{2}\right)
\end{array}
\right)\nonumber\\
& &\left(
\begin{array}{ccc}
l_{1}+\nu_{1} & l_{2}+\nu_{2} & l \\ 
0 & 0 & 0
\end{array}
\right).\nonumber \\
\label{eq:6j}
\end{eqnarray}
The $\left\{...\right\}$ denote the 6j symbol of the Racah-Wigner algebra. It can be computed using 3j symbols as
\begin{eqnarray}
\left\{
\begin{array}{ccc}
a & b & c \\ 
d & e &f 
\end{array}
\right\}
\!\!\!\!&=&\!\!\!\! \sum \left(-1\right)^{d+e+f+\delta+\epsilon+\phi}
\left(\!\!
\begin{array}{ccc}
a & b & c \\ 
\alpha & \beta & \gamma
\end{array}\!\!
\right)
\left(\!\!
\begin{array}{ccc}
a & e & f \\ 
\alpha & \epsilon & -\phi
\end{array}\!\!
\right)\nonumber\\
&&\times\left(\!\!
\begin{array}{ccc}
d & b & f \\ 
-\delta & \beta & \phi
\end{array}\!\!
\right)
\left(\!\!
\begin{array}{ccc}
d & e & c \\ 
\delta & \epsilon & \gamma
\end{array}\!\!
\right),
\label{eqn:sixjdefn}
\end{eqnarray}
where the sum is over all possible values of $\alpha,\beta,\gamma,\delta,\epsilon$ and $\phi$ for which the 3j symbols are defined.

\subsection{Relations between the two basis}

Using the definitions of $\boldsymbol R_{l}^{m}\left(\theta,\varphi\right)$, $%
\boldsymbol S_{l}^{m}\left(\theta,\varphi\right)$, $\boldsymbol
T_{l}^{m}\left(\theta,\varphi\right)$ and the results derived by \cite{Varshalovich1988}, we obtain: 
\begin{equation}
\left\{ 
\begin{array}{l@{\quad}l}
\boldsymbol R_{l}^{m}(\theta,\varphi)=\sqrt{\frac{l}{2l+1}}\boldsymbol
Y_{l,l-1}^{m}\left(\theta,\varphi\right)-\sqrt{\frac{l+1}{2l+1}}\boldsymbol
Y_{l,l+1}^{m}\left(r,\theta\right) &  \\ 
\boldsymbol S_{l}^{m}(\theta,\varphi)=\sqrt{\frac{l}{2l+1}}(l+1)\boldsymbol
Y_{l,l-1}^{m}\left(\theta,\varphi\right)+\sqrt{\frac{l+1}{2l+1}}l\boldsymbol
Y_{l,l+1}^{m}\left(\theta,\varphi\right) &  \\ 
\boldsymbol T_{l}^{m}(\theta,\varphi)=-i\sqrt{l(l+1)}\boldsymbol
Y_{l,l}^{m}(\theta,\varphi) & 
\end{array}
\right.
\label{eq:SVHS}
\end{equation}
and
\begin{equation}
\left\{ 
\begin{array}{l@{\quad}l}
\boldsymbol Y_{l,l-1}^{m}(\theta,\varphi)=\sqrt{\frac{l}{2l+1}}\boldsymbol
R_{l}^{m}\left(\theta,\varphi\right)+\frac{1}{\sqrt{l(2l+1)}}\boldsymbol
S_{l}^{m}\left(\theta,\varphi\right) &  \\ 
\boldsymbol Y_{l,l}^{m}\left(\theta,\varphi\right)=\frac{i}{\sqrt{l\left(l+1\right)}%
}\boldsymbol T_{l}^{m}\left(\theta,\varphi\right) &  \\ 
\boldsymbol Y_{l,l+1}^{m}\left(\theta,\varphi\right)=-\sqrt{\frac{l+1}{2l+1}}\boldsymbol
R_{l}^{m}(\theta,\varphi)+\frac{1}{\sqrt{(l+1)(2l+1)}}\boldsymbol
S_{l}^{m}(\theta,\varphi). & 
\end{array}
\right.
\end{equation}
Therefore, the relations between the coefficients of the projection of a vector ${\boldsymbol X}$ on the $\left({\boldsymbol R},{\boldsymbol S},{\boldsymbol T}\right)$ and on the $\left\{{\boldsymbol Y}_{l,l+\nu}^{m}\right\}_{\nu=\left\{-1,0,1\right\}}$ basis are given by: 
\begin{equation}
\left\{ 
\begin{array}{l@{\quad}l}
u_{m}^{l}(r)=\frac{1}{\sqrt{2l+1}}\left[\sqrt{l}X_{l,l-1}^{m}(r)-\sqrt{l+1}%
X_{l,l+1}^{m}(r)\right] &  \\ 
v_{m}^{l}(r)=\frac{1}{\sqrt{2l+1}}\left[\frac{1}{\sqrt{l}}X_{l,l-1}^{m}(r)+%
\frac{1}{\sqrt{l+1}}X_{l,l+1}^{m}(r)\right] &  \\ 
w_{m}^{l}(r)=\frac{i}{\sqrt{l(l+1)}}X_{l,l}^{m}(r) & 
\end{array}
\right.
\end{equation}
and
\begin{equation}
\left\{ 
\begin{array}{l@{\quad}l}
X_{l,l-1}^{m}\left(r\right)=\sqrt{\frac{l}{2l+1}}\left(u_{m}^{l}\left(r%
\right)+(l+1)v_{m}^{l}\left(r\right)\right) &  \\ 
X_{l,l}^{m}\left(r\right)=-i\sqrt{l\left(l+1\right)}w_{m}^{l}\left(r\right)
&  \\ 
X_{l,l+1}^{m}\left(r\right)=\sqrt{\frac{l+1}{2l+1}}\left(-u_{m}^{l}\left(r%
\right)+l v_{m}^{l}\left(r\right)\right). & 
\end{array}
\right.  \label{chb2}
\end{equation}
This allows us to obtain the expansion on spherical harmonics of the scalar product of two general vectors initially projected on the $\left({\boldsymbol R},{\boldsymbol S},{\boldsymbol T}\right)$ basis using the SVH basis where this product can be analytically computed using the Racah-Wigner algebra.

\section{Matrix elements in the dipolar case}
\label{Sec:AppendixDipolar}

In the case of dipolar high-order g modes, the matrix elements $M_{m,m'}$ defined in Eq. (\ref{eq:MatrixLorentz}) simplify and can be expressed analytically as $\omega_{\rm B}$ and $a$ using the Racah-Wigner algebra. We obtain 
\begin{eqnarray}
\lefteqn{M_{1,1}=M_{-1,1}=}\nonumber\\
&&\frac{1}{2\mu_0\omega_0 I}\frac{3}{4}\frac{1}{2\pi}\int_{r_i}^{r_o}\left[\partial_r\left(r\xi_{h}\right)\right]^2\int_{\Omega=4\pi}B_{r}^{2}\left(1+\cos^{2}\theta\right){\rm d}\Omega{\rm d}r,\nonumber\\
\end{eqnarray}
where ${\rm d}\Omega=\sin\theta{\rm d}\theta{\rm d}\varphi$ and $\omega_{0;n,l}$ and $\xi_{h;n,l}$ have been simplified onto $\omega_{0}$ and $\xi_{h}$, respectively. By expanding $1+\cos^{2}\theta$ on spherical harmonics, using the definition of $B_{r}$ given in Eq. (\ref{eq:Br}) and the coupling integrals $\left(\mathcal J\right)$ provided in Eq. (\ref{eq:3j}), and assuming the JWKB approximation, we derive its final expression:
\begin{eqnarray}
\lefteqn{M_{1,1}=M_{-1,-1}=}\nonumber\\
&&\frac{1}{2\mu_0\omega_0^3}\frac{1}{2\pi}\sqrt{\pi}{\mathcal I}\sum_{l_1=1}^{L}\sum_{m_1=-l_1}^{l_1}\sum_{l_2=1}^{L}\sum_{m_2=-l_2}^{l_2}\nonumber\\
&&\int_{r_i}^{r_o} K\left(r\right)b_{m_1}^{l_1}\left(r\right)b_{m_2}^{l_2}\left(r\right)\left(2{\mathcal J}_{l_1,m_1,l_2,m_2}^{0,0}+\frac{1}{\sqrt{ 5}}{\mathcal J}_{l_1,m_1,l_2,m_2}^{2,0}\right){\rm d}r,\nonumber\\
\end{eqnarray}
where $K\left(r\right)$ and ${\mathcal I}$ have been defined in Eq. (\ref{eq:kernel}).

Applying the same method to the other matrix elements, we get:
\begin{eqnarray}
\lefteqn{M_{0,0}=\frac{1}{2\mu_0\omega_0 I}\frac{3}{2}\frac{1}{2\pi}\int_{r_i}^{r_o}\left[\partial_r\left(r\xi_{h}\right)\right]^2\int_{\Omega=4\pi}B_{r}^{2}\left(1-\cos^{2}\theta\right){\rm d}\Omega{\rm d}r,}\nonumber\\
&&=\frac{1}{2\mu_0\omega_0^3}\frac{1}{2\pi}2\sqrt{\pi}{\mathcal I}\sum_{l_1=1}^{L}\sum_{m_1=-l_1}^{l_1}\sum_{l_2=1}^{L}\sum_{m_2=-l_2}^{l_2}\nonumber\\
&&\int_{r_i}^{r_o} K\left(r\right)b_{m_1}^{l_1}\left(r\right)b_{m_2}^{l_2}\left(r\right)\left({\mathcal J}_{l_1,m_1,l_2,m_2}^{0,0}-\frac{1}{\sqrt{ 5}}{\mathcal J}_{l_1,m_1,l_2,m_2}^{2,0}\right){\rm d}r,\nonumber\\
\end{eqnarray}
\begin{eqnarray}
\lefteqn{M_{0,1}=-M_{-1,0}}\nonumber\\
&&=\frac{1}{2\mu_0\omega_0 I}\frac{3}{2\sqrt{2}}\frac{1}{2\pi}\\
&&\int_{r_i}^{r_o}\left[\partial_r\left(r\xi_{h}\right)\right]^2\int_{\Omega=4\pi}B_{r}^{2}\sin\theta\cos\theta\exp\left[i\varphi\right]{\rm d}\Omega{\rm d}r,\nonumber\\
&&=-\frac{1}{2\mu_0\omega_0^3}\frac{1}{2\pi}\sqrt{\frac{3\pi}{5}}{\mathcal I}\sum_{l_1=1}^{L}\sum_{m_1=-l_1}^{l_1}\sum_{l_2=1}^{L}\sum_{m_2=-l_2}^{l_2}{\mathcal J}_{l_1,m_1,l_2,m_2}^{2,1}\nonumber\\
&&\int_{r_i}^{r_o} K\left(r\right)b_{m_1}^{l_1}\left(r\right)b_{m_2}^{l_2}\left(r\right){\rm d}r,\nonumber\\
\end{eqnarray}
and
\begin{eqnarray}
\lefteqn{M_{-1,1}=M_{1,-1}^{*}=}\nonumber\\
&&\frac{1}{2\mu_0\omega_0 I}\frac{3}{4}\frac{1}{2\pi}\int_{r_i}^{r_o}\left[\partial_r\left(r\xi_{h}\right)\right]^2\int_{\Omega=4\pi}B_{r}^{2}\sin^{2}\theta\exp\left[i2\varphi\right]{\rm d}\Omega{\rm d}r,\nonumber\\
&&=\frac{1}{2\mu_0\omega_0^3}\frac{1}{2\pi}\sqrt{\frac{6\pi}{5}}{\mathcal I}\sum_{l_1=1}^{L}\sum_{m_1=-l_1}^{l_1}\sum_{l_2=1}^{L}\sum_{m_2=-l_2}^{l_2}{\mathcal J}_{l_1,m_1,l_2,m_2}^{2,2}\nonumber\\
&&\int_{r_i}^{r_o} K\left(r\right)b_{m_1}^{l_1}\left(r\right)b_{m_2}^{l_2}\left(r\right){\rm d}r,\nonumber\\
\end{eqnarray}\\
while $M_{1,0}=-M_{0,-1}=M_{0,1}^{*}$.

If we define $b=\zeta {\omega}_{\rm B}/\omega_{\rm R}$, where $\omega_{\rm B}$ and $\omega_{\rm R}$ are given in Eqs. (\ref{eq:omegaB}) \& (\ref{eq:rotation}), respectively, which evaluates the relative strength of magnetic and rotation terms, and $c=M_{0,1}/\omega_{\rm B}$ and $d=M_{-1,1}/\omega_{\rm B}$, which are related to the non-axisymmetry of $B_r^2$, the matrix $\zeta{\boldsymbol M}+{\boldsymbol R}$ simplifies onto:
\begin{equation}
    \zeta{\boldsymbol M}+{\boldsymbol R}=\omega_{\rm R}
    \begin{bmatrix}
     b\left(1+\dfrac{a}{2}\right)-1& -bc & bd  \\
     -bc^* & b(1-a) & bc \\
     bd^* & bc^* & b\left(1+\dfrac{a}{2}\right)+1
    \end{bmatrix}.
\end{equation}

\end{appendix}

\end{document}